\documentclass{appolb}
\usepackage{graphicx}

\begin{document}
\title{The chiral long-range two-pion exchange electromagnetic currents in radiative nucleon-deuteron capture
}
\author{R.Skibi\'nski, J.Golak, D.Rozp\c edzik, K.Topolnicki, H.Wita{\l}a
\address{M.Smoluchowski Institute of Physics, Jagiellonian University, PL-30348, Krak\'ow, Poland}
}
\maketitle
\begin{abstract}
The nucleon-deuteron radiative capture process is investigated using the chiral nuclear 
potentials and the electromagnetic currents developed by the Bochum-Bonn group.
While the strong interaction is taken up to the next-to-next-to-leading order, the 
electromagnetic current consists of a single nucleon current, the leading one-pion exchange one
and is supplemented by contributions from
the long-range two-pion exchange current at next-to-leading-order.
The theoretical predictions for the cross sections as well as analyzing powers 
show strong dependence on the values of regularization parameters. Only small effects of the
three-nucleon force and the long-range two-pion exchange current are observed. 
The dependence on the choice of regularization parameters results in 
a big theoretical uncertainty and clearly points to the necessity to include corrections from higher 
orders of the chiral expansion both for the nuclear forces and currents. 
 
\end{abstract}
  
\section{Introduction}
In the recent decade the chiral Effective Field Theory ($\chi$EFT) has proven its predictive power 
in low energy nuclear physics. In the nucleon/pion sector it was successfully applied 
to such processes as 
pion-nucleon scattering, nucleon-nucleon scattering, three-nucleon reactions and 
to the analysis of heavier systems. For a review on the $\chi$EFT and its recent applications 
see e.g.\cite{Epelbaum_review, Epelbaum_review2, Machleidt_review}.
Among many theoretical advantages of the $\chi$EFT the
consistency of the derived two- and three-body forces and interrelated single and many-body electromagnetic and week 
currents should be mentioned. This predisposes the $\chi$EFT to be used 
in studies of electromagnetic processes. 

On the other hand, the inclusion of the electromagnetic current can be done also in an approximate way 
by means of the Siegert theorem~\cite{Siegert}. Both approaches lead to similar results 
for proton-deuteron (pd) capture and for photodisintegration processes~\cite{Golak_pdcap,skib2b}.
In this work we extend the results of \cite{Skibinski_APP} by using explicitly the chiral electromagnetic current
instead of the Siegert approach. We follow here the path charted in \cite{Rozpedzik} 
and use the single nucleon current, the leading one-pion exchange current and the next-to-leading order (NLO) contributions to the 
long-range part of
the two-pion (2$\pi$) exchange current.
In~\cite{Rozpedzik} the effects of the 2$\pi$-exchange currents were explored mainly in 
the deuteron photodisintegration reaction. It was found that the long-range 2$\pi$-exchange current 
operator plays an important role for that process.

In the next section we briefly present our formalism for the nucleon-deuteron radiative capture
and give details on the electromagnetic current. In Sec.~\ref{Results} we present predictions for 
pd-capture at two laboratory energies of the incoming deuteron: 17.5 MeV and 137 MeV. 
These deuteron energies correspond to photon energies 11.3 and 50.4 MeV in the centre-of-mass system, respectively.
We conclude in Sec.~\ref{Summary}.

\section{Theoretical description of the radiative Nd-capture reaction}
The radiative pd-capture process $p+d \rightarrow \gamma + ^3$He is connected to the two-body photodisintegration of $^3$He via 
the time reversal symmetry. Thus we obtain the nuclear matrix elements $N^{rad}_{\mu}$ for the radiative 
nucleon-deuteron (Nd) capture reaction from the matrix elements $N^{Nd}_{\mu} \equiv \langle \Psi_{Nd}^{(-)} \mid j_{\mu} \mid \Psi_b \rangle$ 
for the photodisintegration process.
The $N^{Nd}_{\mu}$ can be expressed as
\begin{eqnarray}
N_\mu^{Nd} =
\langle \phi_{1} \mid  ( 1 + P ) \mid
j_\mu \mid \Psi_b \rangle +
\langle \phi_{1} \mid  P \mid \tilde U \rangle ,
\label{eq:Nnew}
\end{eqnarray}
where $\mid \phi_1 \rangle$ is a product of the deuteron state
and a momentum eigenstate of the spectator nucleon.
Further, the $\mid \Psi_b \rangle$ is the initial three nucleon (3N) bound state,
$j_{\mu}$ is the $\mu$-component of the electromagnetic current
operator and $P\equiv P_{12}P_{23} + P_{13}P_{23}$ is the permutation operator
with two-body permutations $P_{ij}$ interchanging  
the i-th and j-th nucleons.

The auxiliary state $\mid \tilde U \rangle$ fulfills
the Faddeev-like equation~\cite{raport2005}
\begin{eqnarray}
\mid \tilde U \rangle  &=&
\left( t G_0 + \frac12 ( 1 + P ) V_4^{(1)} G_0 ( t G_0 +1) \right) ( 1 + P ) j_\mu \mid \Psi_b \rangle \nonumber \\
&+& \left( t G_0 P + \frac12 ( 1 + P ) V_4^{(1)} G_0 ( t G_0 +1) P \right)
\mid \tilde U \rangle ,
\label{eq:Utilde}
\end{eqnarray}
where $V_4^{(1)}$ is a part of the 3N force which is
symmetrical under the exchange of nucleons 2 and 3, $G_0$ is
the free 3N propagator,  and $t$ is the
t-matrix connected to the nucleon-nucleon (NN) interaction via the Lippmann-Schwinger equation. 

We solve Eq.(\ref{eq:Utilde}) in the mo\-men\-tum-spa\-ce in the partial wave scheme.
More details on our formalism and numerical performance can be 
found in ~\cite{skib2b, raport2005}. 
As the nuclear interaction we use the NN and 3N potentials at the next-to-next-to-leading order (N$^2$LO) developed by 
E.Epelbaum and collaborators in~\cite{epelbaum2005} and~\cite{epelbaum3NFatN2LO}, respectively.
These forces depend on two regularization parameters $(\Lambda,\tilde{\Lambda})$ used in the regularization of
the Lippmann-Schwinger equation and three-nucleon force and in the spectral function regularization 
of 2$\pi$ exchanges in the chiral NN interaction~\cite{Epelbaum_lambda}.
The values of regularization parameters $(\Lambda,\tilde{\Lambda})$ are taken as in~\cite{epelbaum3NFatN2LO} and are
listed below. 
We neglect the Coulomb interaction between protons in the initial proton-deuteron system.
The nuclear electromagnetic current has been developed within the same approach 
and presented in~\cite{Kolling1,Kolling2} and used in~\cite{Rozpedzik}. 
Here we use the same current operator as in~\cite{Rozpedzik}, where its structure  
and partial wave decomposition is discussed in detail.

\section{Results}
\label{Results}
\begin{figure}[htb]
\centerline{%
\includegraphics[width=12.5cm,clip=true]{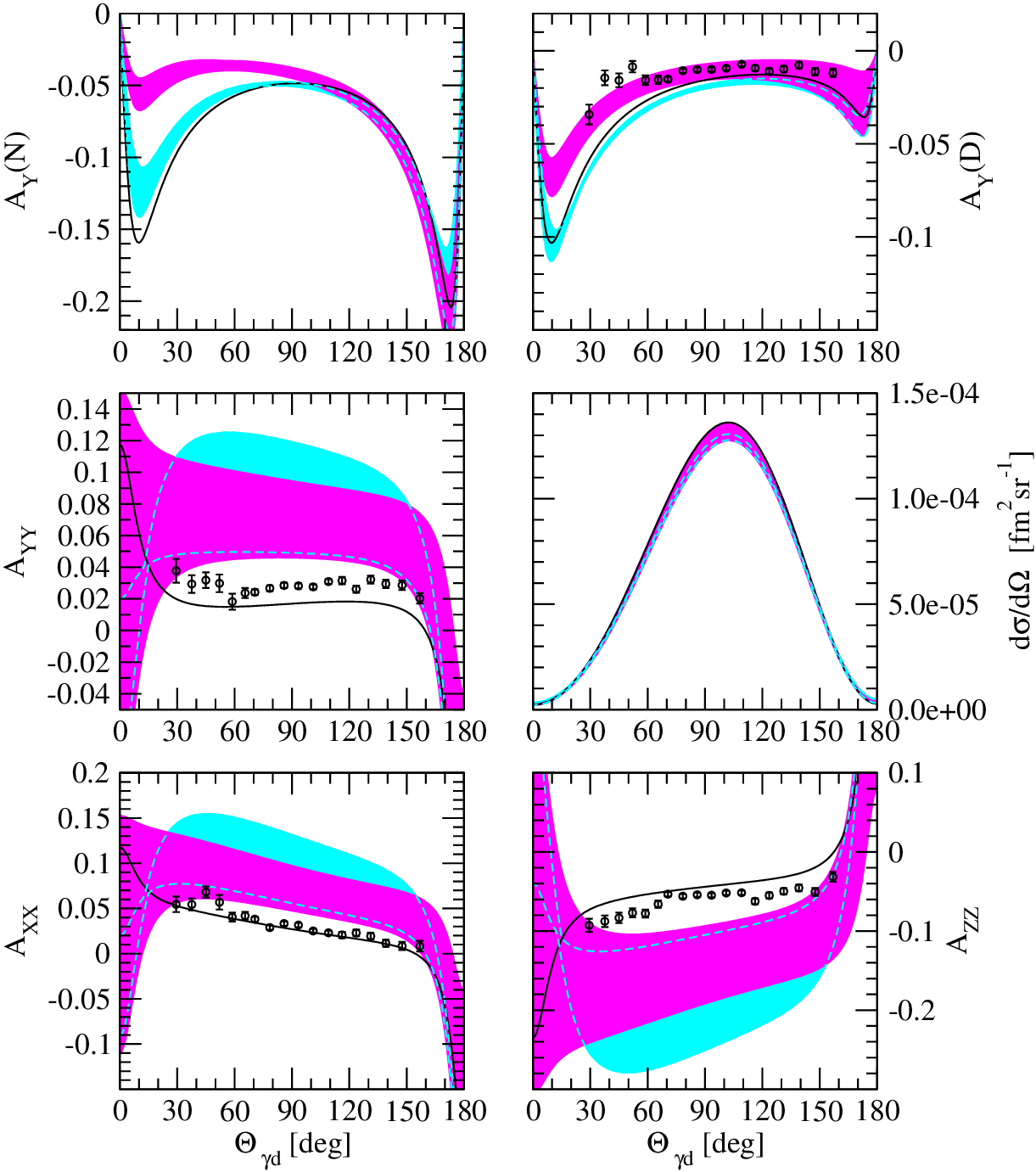}}
\caption{The differential cross section and the analyzing powers for the proton-deuteron radiative capture
at E$_d$=17.5 MeV. The light (blue) band corresponds to predictions which base on a
single nucleon and one-pion exchange currents. The dark (magenta) band corresponds to predictions which base on
the same currents supplemented by the long-range 2$\pi$-exchange currents at NLO (see text).
The dashed line shows the borders of the lighter band.
The solid black line is for the AV18+Urbana IX and the standard meson exchange currents (see text) predictions.
Experimental data points are taken from \cite{Sagara.1,Sagara.2}.}
\label{fig1}
\end{figure}

\begin{figure}[htb]
\centerline{%
\includegraphics[width=12.5cm,clip=true]{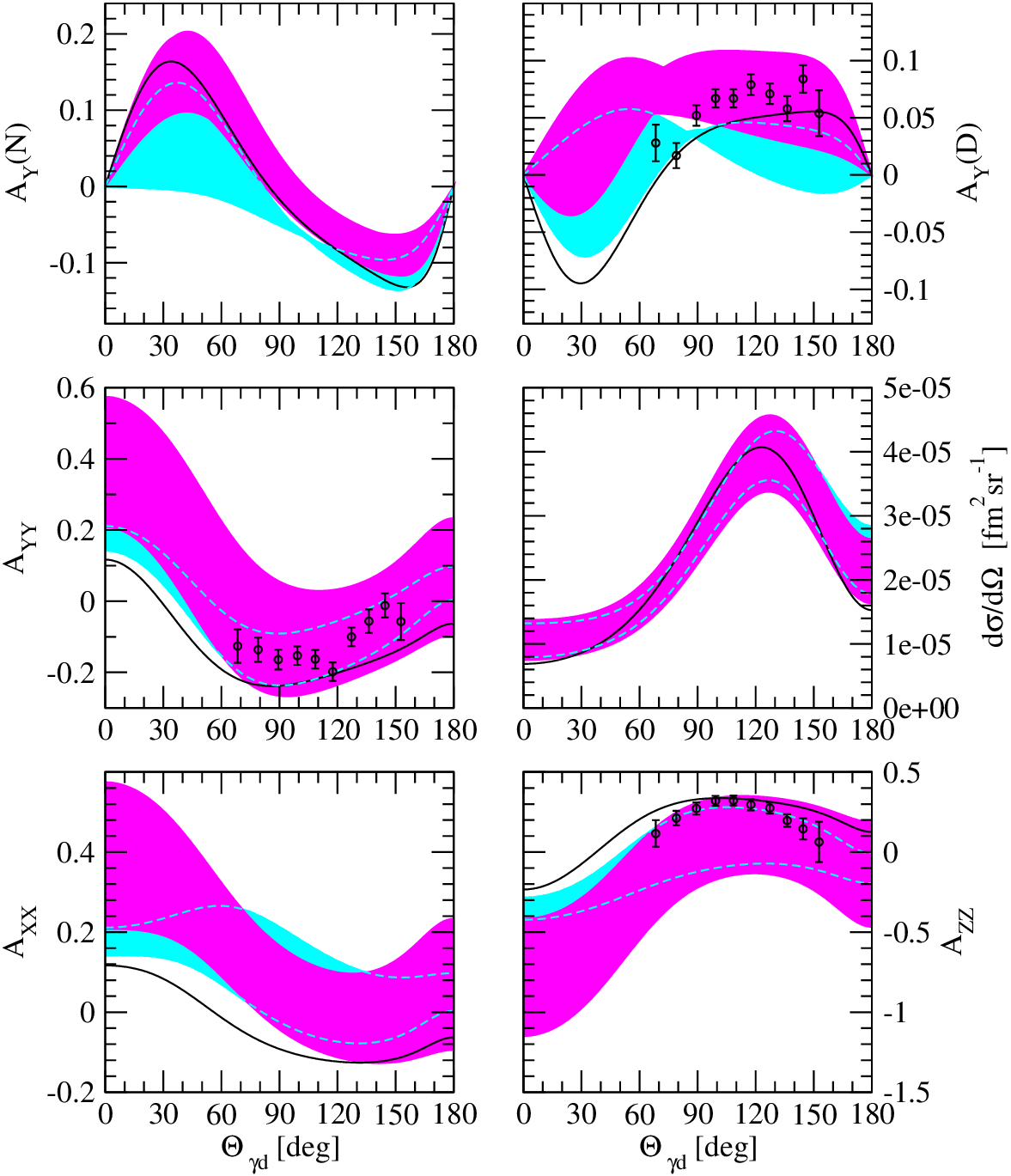}}
\caption{The same as in Fig.~\ref{fig1} but for the deuteron laboratory energy E$_d$=137 MeV.
Experimental data points at E$_d$=133 MeV are taken from~\cite{KVIdata}}
\label{fig2}
\end{figure}

In Figs.~\ref{fig1}-\ref{fig2} we show predictions obtained with a complete nuclear potential 
(NN + 3N) at N$^2$LO and two models of the electromagnetic current.
The light blue band comprises predictions for different values of the regularization 
parameters $(\Lambda,\tilde{\Lambda})$ 
obtained with
the single nuclear current and the leading one-pion exchange current. In predictions which contribute to the 
dark magenta band also long-range corrections to the 2$\pi$-exchange current 
are included. In addition Fig.~\ref{fig1} also shows the prediction based on the AV18~\cite{wiringa_AV18} 
NN force combined with
the Urbana IX 3N force~\cite{UrbIX}.
In this case the electromagnetic current is a sum of the single nucleon current and two most 
important contributions to the meson exchange current: the so called $\pi$-like and $\rho$-like terms~\cite{Golak_pdcap}.
At the lower energy E$_d$=17.5 MeV the differential cross section and both (the nucleon and the deuteron)
vector analyzing powers are rather insensitive to the values of $(\Lambda,\tilde{\Lambda})$, what leads to a relatively narrow bands.
The predictions for tensor analyzing powers are much more sensitive to the values of the regularization parameters.
The quality of the data~\cite{Sagara.1,Sagara.2} description is similar for both models of
electromagnetic current and only slight improvement is seen after inclusion of the NLO contributions.
At the higher energy E$_d$=137 MeV, all bands are much wider and no meaningful differences in predictions
within two models of electromagnetic current are observed.   

\begin{figure}[htb]
\centerline{%
\includegraphics[width=12.5cm,clip=true]{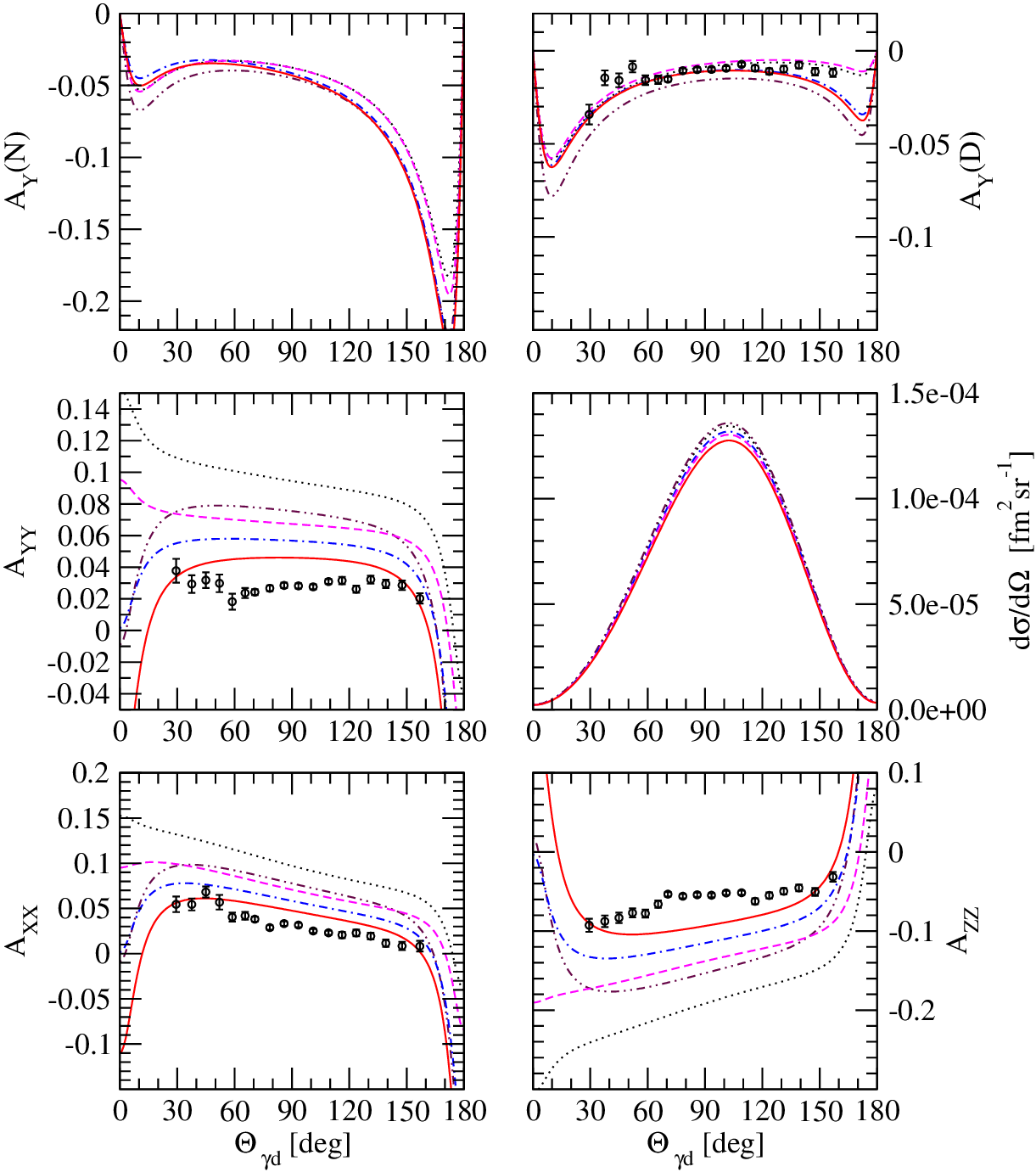}}
\caption{The same as in Fig.~\ref{fig1} but now only contributions to the dark (magenta) band
are shown explicitly. Values of the $(\Lambda,\tilde{\Lambda})$ parameters are given in the text.}
\label{fig3}
\end{figure}

In Fig.~\ref{fig3} we show in more detail the dependence of the predictions on the value of
regularization parameters. The presented lines are the same that build the dark, magenta band
in Fig.~\ref{fig1}. 
The values of $(\Lambda,\tilde{\Lambda})$ are, in MeV: (450,500),(600,500),(550,600),(450,700),(600,700) for the
black dotted, brown double-dot-dashed, blue dot-dashed, magenta dashed and red solid curves, respectively. 
The strongest dependence is seen for the tensor analyzing powers. Note, that predictions based on the smallest values 
of $\Lambda$ behave qualitatively similar to the
reference predictions based on the AV18+Urbana IX model, shown in Fig.~\ref{fig1}. This is in agreement with recent
findings in elastic Nd scattering at N$^3$LO where also the smallest values of $\Lambda$ are favoured~\cite{Golak_LENPIC}.

\begin{figure}[htb]
\centerline{%
\includegraphics[width=12.5cm,clip=true]{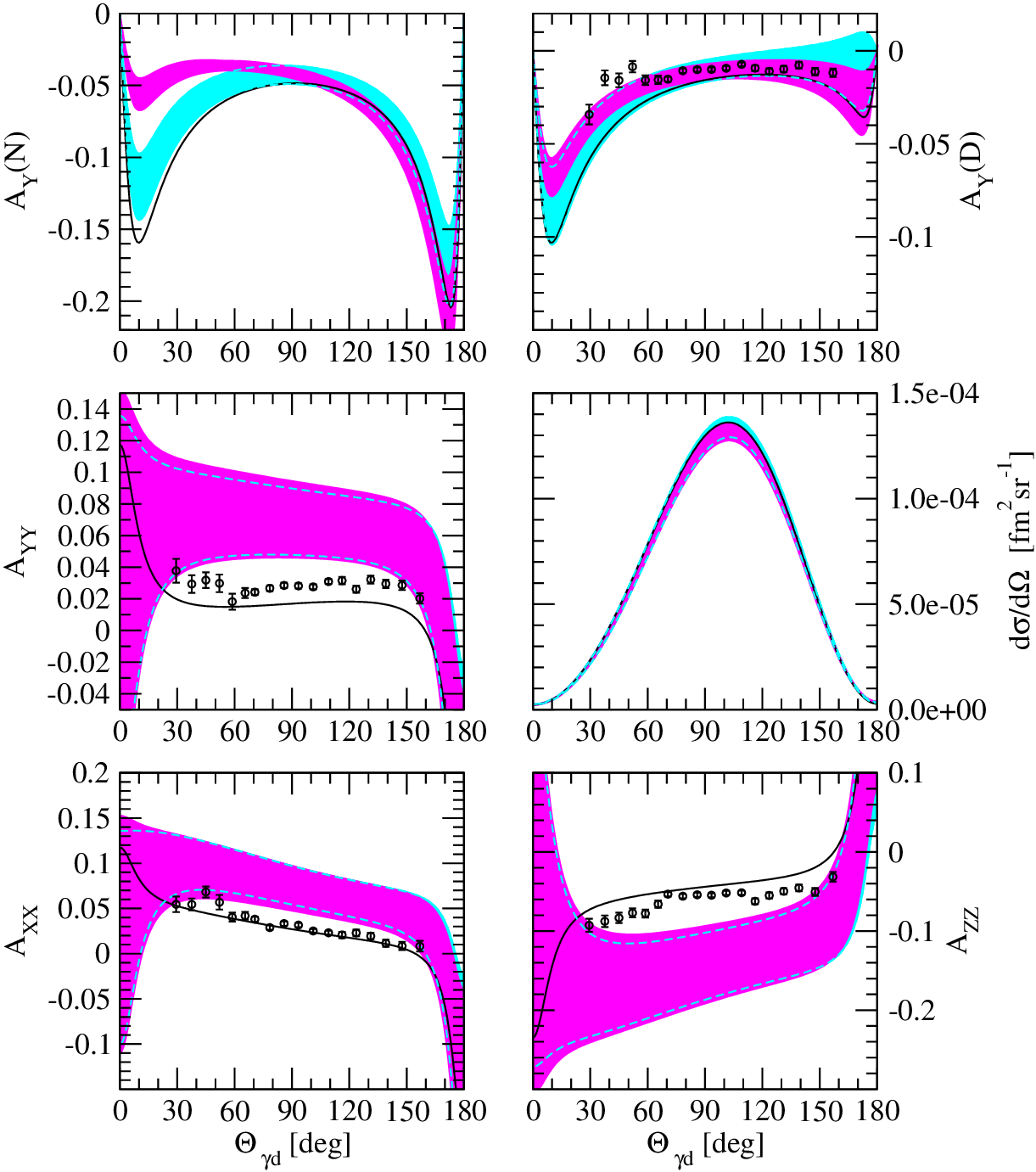}}
\caption{The same as in Fig.~\ref{fig1} but now for predictions based on 
the NN N$^2$LO interaction only (light (blue) band) and the full NN+3N Hamiltonian at N$^2$LO (dark (magenta) band).
The solid line and experimental data points are the same as in Fig.~\ref{fig1}.}
\label{fig4}
\end{figure}

\begin{figure}[htb]
\centerline{%
\includegraphics[width=12.5cm,clip=true]{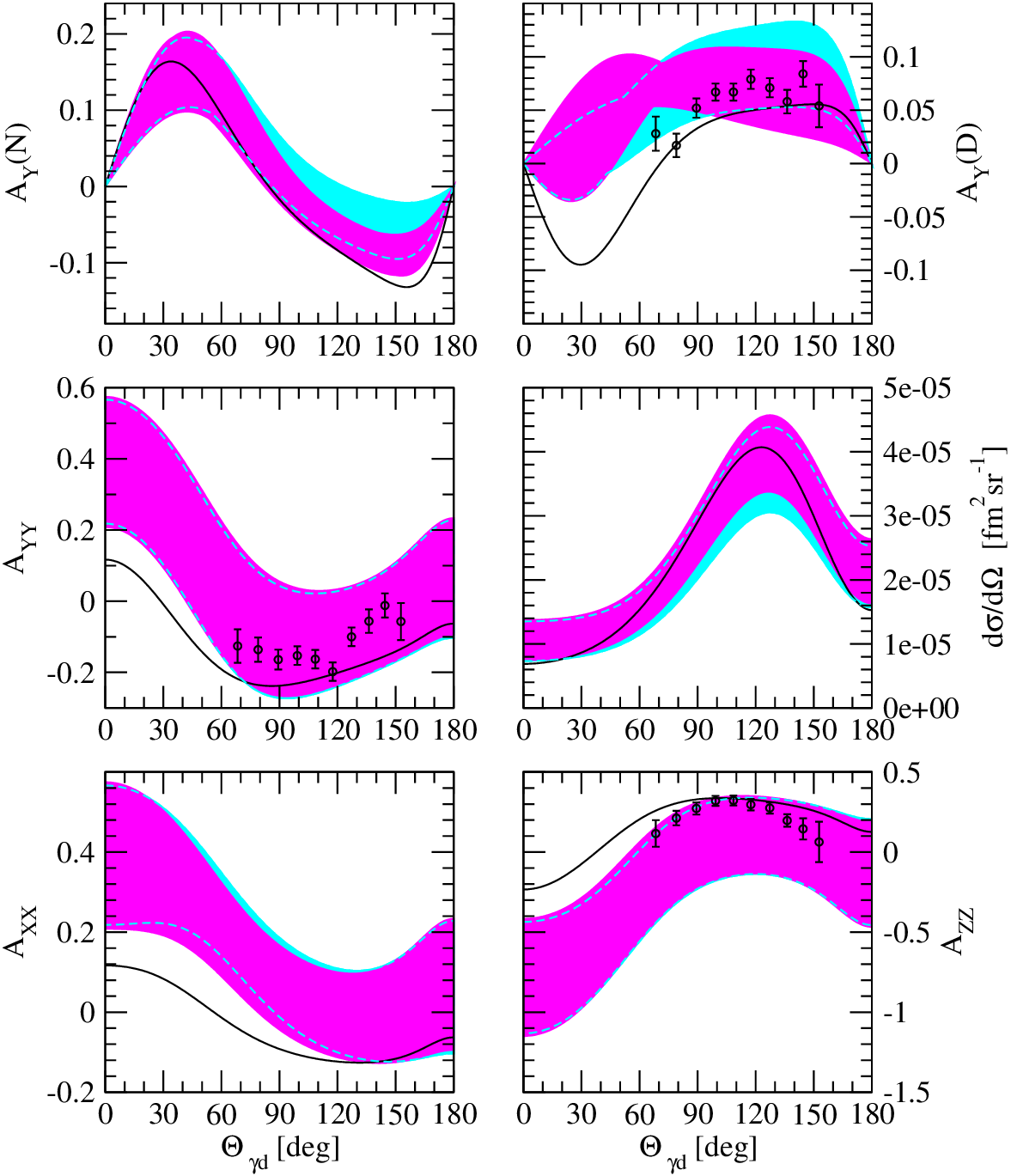}}
\caption{The same as in Fig.~\ref{fig4} but for the deuteron laboratory energy E$_d$=137 MeV.
Experimental data points from~\cite{KVIdata} are taken at E$_d$=133 MeV. }
\label{fig5}
\end{figure}

Finally, we test the sensitivity of the Nd-capture observables to the 
3N force. In Fig.~\ref{fig4} we compare predictions based on 
the NN interaction only (light blue band) with ones based on the full NN+3N Hamiltonian
(dark magenta band) for E$_d$=17.5 MeV. The electromagnetic current operator is again taken as a sum 
of the single nucleon current, the one-pion exchange current and the long-range 2$\pi$-exchange one. 
At this energy only predictions for the differential cross section and vector analyzing powers 
change after inclusion of the 3N force. However this change is not big and comparable with 
the bands' width. For tensor analyzing powers the 3N force effects are totally 
hidden by the dependence on the values of the regularization parameters.
The latter is true also at E$_d$=137 MeV (Fig.~\ref{fig5}) where, additionally, the  
3N force effects are seen neither for the
vector analyzing powers nor for the cross section.
Again, for these dynamical models the bands are too broad to be useful in a study of the 3N force effects.

\section{Summary}
\label{Summary}

We have applied the chiral potential at N$^2$LO combined with the electromagnetic current
taken as a sum of the single nucleon current, the one-pion exchange one and the long-range NLO part 
of the 2$\pi$-exchange current to study the differential cross section and the analyzing powers 
for the Nd radiative capture at deuteron energies 17.5 and 137 MeV.

For most observables the effects of long-range 2$\pi$-exchange current at NLO 
as well as the effects of the
3N interaction at N$^2$LO  are small compared to the uncertainty coming from different
values of the regularization parameters. This clearly shows that a study of the Nd-capture reaction
by means of $\chi$EFT requires consistent currents and interactions at higher orders of
the chiral expansion.
In addition, more constraints should be put on the values of the regularization parameters.

Recently it was pointed that the dependence of the predictions for the Nd 
elastic scattering on the regularization parameters obtained within
the ($\chi$EFT) at N$^3$LO also 
leads to unacceptably large theoretical uncertainties~\cite{WitalaJPG, Golak_LENPIC}.
It is expected, that a new regularization scheme will cure such a situation~\cite{Epelbaum_private}.
Once it is available, it will be interesting to study
the Nd-capture and other electromagnetic processes again. 
Such an investigation is planed in the future.

\section{Acknowledgements}
The authors would like to thank Prof. E.Epelbaum and Dr. S.K\"olling 
for their help in the numerical realization of chiral interactions and current operators.
This study 
was supported by the Polish National Science Center under Grants No.DEC-
2013/10/M/ST2/00420 and DEC2013/11/N/ST2/03733. 
We acknowledge support by the Foundation for Polish Science - MPD program, co-financed 
by the European Union within the European Regional Development Fund.
The numerical calculations have been performed on the supercomputer
clusters of the JSC, Julich, Germany.

\end{document}